\documentclass[conference]{IEEEtran}
\IEEEoverridecommandlockouts
\input{config.sty}
\usepackage[show]{chato-notes}
\usepackage{booktabs}
\usepackage[cm]{fullpage}
\usepackage{amsmath}
\usepackage{caption}
\IEEEoverridecommandlockouts

\begin{document}

\newcommand{\nic}[1]{\textcolor{blue}{#1}}
\newcommand{\albe}[1]{\textcolor{red}{#1}}
\newcommand{\pietro}[1]{\textcolor{green}{#1}}

\bstctlcite{IEEEexample:BSTcontrol}

\title{A Federated Channel Modeling System using Generative Neural Networks \thanks{This work was partially supported by the European Union under the Italian National Recovery and Resilience Plan (PNRR) of NextGenerationEU, partnership on "Telecommunications of the Future" (PE00000001 - program "RESTART") and by the HORIZON-CL4-2021-SPACE-01 project "5G+ evoluTion to mutioRbitAl multibaNd neTwORks" (TRANTOR) No. 101081983}
}

\author{
\IEEEauthorblockN{
Saira Bano\IEEEauthorrefmark{1}\textsuperscript{,}\IEEEauthorrefmark{2},
Pietro Cassarà\IEEEauthorrefmark{2},
Nicola Tonellotto\IEEEauthorrefmark{1},
Alberto Gotta\IEEEauthorrefmark{2}\\
 }
\IEEEauthorblockA{
\IEEEauthorrefmark{1}Department of Information Engineering, University of Pisa, Pisa, Italy \\
\IEEEauthorrefmark{2}Information Science and Technology Institute "A. Faedo", National Research Council, Pisa, Italy \\
\texttt{saira.bano@phd.unipi.it, pietro.cassara@isti.cnr.it}\\
\texttt{nicola.tonellotto@unipi.it, alberto.gotta@isti.cnr.it}
}
\par}

\maketitle
\begin{abstract}
The paper proposes a data-driven approach to air-to-ground channel estimation in a millimeter-wave wireless network on an unmanned aerial vehicle. Unlike traditional centralized learning methods that are specific to certain geographical areas and inappropriate for others, we propose a generalized model that uses Federated Learning (FL) for channel estimation and can predict the air-to-ground path loss between a low-altitude platform and a terrestrial terminal. To this end, our proposed FL-based Generative Adversarial Network (FL-GAN) is designed to function as a generative data model that can learn different types of data distributions and generate realistic patterns from the same distributions without requiring prior data analysis before the training phase. To evaluate the effectiveness of the proposed model, we evaluate its performance using Kullback-Leibler divergence (KL), and Wasserstein distance between the synthetic data distribution generated by the model and the actual data distribution. We also compare the proposed technique with other generative models, such as FL-Variational Autoencoder (FL-VAE) and stand-alone VAE and GAN models. The results of the study show that the synthetic data generated by FL-GAN has the highest similarity in distribution with the real data. This shows the effectiveness of the proposed approach in generating data-driven channel models that can be used in different regions.
\end{abstract}


\section{Introduction}
\label{sec:introduction}
Non-terrestrial networks (NTNs), such as near-earth satellite constellations (LEO), high-altitude platforms, and unmanned aerial vehicles (UAVs) have traditionally been used for disaster management and remote sensing \cite{azari2022evolution}. However, they are now being seen as promising technologies for providing ubiquitous connectivity in the future generation of the Internet \cite{giordani2020non}.
Such radio access networks (RAN), operating in the millimeter wave (mmWave) range, are very promising, providing global coverage and high capacity for reliable and efficient communications services \cite{giordani2020satellite}.
The 3rd Generation Partnership Project (3GPP), 
has also recognized the potential of mmWave technology to support satellite communications. 


Accurate statistical channel models are essential to characterize the mmWave link and to determine the underlying channel parameters to improve the transmission performance of wireless communication systems. Extensive research has been conducted to develop effective methods for accurate channel modeling, such as the mathematical propagation model proposed in \cite{Shakhatreh2021ModelingGP} for estimating ground-to-air path loss between wireless devices and low-altitude platforms using mmWave frequency bands. Furthermore, deterministic channel models, such as ray-tracing techniques, as well as stochastic channel models are commonly used and require extensive technical knowledge and expertise for analyzing measurement data to estimate a comprehensive set of different channel parameters \cite{zhu2013cost}. However, building statistical channel models to determine the underlying channel parameters that accurately capture the delay, direction, and path gains of individual links is difficult, especially in the mmWave domain. 

Machine Learning (ML) techniques, such as Neural Networks (NNs), can be used to develop statistical channel models that overcome the limitations of conventional channel modeling systems \cite{stocker1993neural}. However, these models result in channel parameters that are site-specific and may not be generally applicable. In this regard, generative NNs, which have proven to be very successful in modeling images and text, provide a suitable approach to data-driven channel modeling and can accurately represent complex environments. Initial research has explored the use of generative NNs for site-specific wireless channels. For example, in \cite{geraci}, the authors proposed generative networks to model channel parameters and trained five different models for five different cities. In contrast, our main goal is to develop a general model that can be used for all participating cities, considering an acceptable model performance for each of these different locations. 



To this end, we propose a location-agnostic statistical channel propagation model based on Federated Learning (FL) that focuses on predicting the path loss component between a UAV and terrestrial nodes in mmWave communication networks. FL is a paradigm developed by Google that aims to build ML models with distributed datasets across multiple devices while maintaining privacy \cite{mcmahan2017communication}.
Participating users communicate parameters or gradients to a central server, which updates and distributes a global model without access to user data \cite{tonellotto2021neural,cassara2022federated,bano2022federated}

@inproceedings{bano2022federated,}. However, in this work, we used the FL frameworks as distributed training engines to train our models on different datasets and develop the generalized channel model
using Variational Autoencoder (VAE) and Conditional Generative Adversarial Network (CGAN) architectures, i.e. FL-VAE and FL-GAN. In our study, 
 we rely on the statistical characteristics of the urban environment of the target area collected through ray tracing simulations to train the models. The performance of the proposed approach is determined using various statistical parameters.

The remainder of the paper is organised as follows. Section \ref{sec:system} discusses the system model, while sections \ref{sec:federatedVAE}
and \ref{sec:federatedcgan} present the federated VAE and GAN approaches for channel modeling, respectively. Section \ref{sec:simulationresults} shows the experimental evaluation performed. Finally, Section \ref{sec:conclusions} draws conclusions.

\section{System Model}
\label{sec:system}
\begin{figure}[t]
    \centering
    \includegraphics[width=\columnwidth]{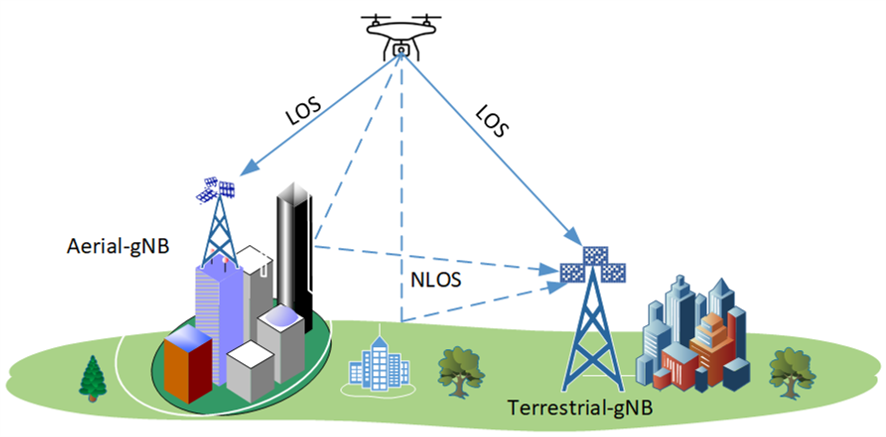}
    \caption{System Model}
    \label{fig:systemmodel}
\end{figure}

In this work, we focus on channel parameter modeling with the main focus on the path loss component connecting UAVs to cellular base stations on the ground, i.e. gNB. We propose a distributed training approach using FL for channel model estimation with two generative NNs. For modelling purposes, we assume that the UAVs act as transmitters and the ground base stations act as receivers, but the roles can be reversed. To model the air-to-ground channel, we assume two ground gNBs, one terrestrial and the other aerial, as in \cite{geraci2}. The aerial gNBs serve as dedicated stations (mounted on rooftops and tilted upward), while the terrestrial gNBs are for ground users (mounted at street level), as shown in Figure \ref{fig:systemmodel}. In addition, we assume three link states between the transmitter and receiver, including Line of Sight (LOS), Non-LOS (NLOS), and no link (i.e., no paths are available). However, when modelling path loss between UAVs and gNBs, we mainly focus on NLOS paths since for LOS, path loss can be calculated using Friis' law \cite{heath2018foundations}.

We adopt the channel parameters estimated with the raytracer package by \cite{geraci} as a benchmark dataset for our investigation. The raytracer simulations estimate the channel parameters, including path losses, azimuth and elevation angles of arrival and departure, and propagation delays. 
According to the dataset, there is a total of 20 paths per link and six parameters per path, resulting in 120 parameters per link with a maximum path loss of 200 dB \cite{geraci}.
The dataset consists of channel parameters for different cities estimated by using the ray-tracer package. 
Using this dataset, we train the generative models for each city in a decentralized manner. These standalone models can learn the channel representation of a UAV's local dataset in a given region but may have biases and be applicable only in a limited spatial domain. Therefore, a general model that is not tied to a specific environment is essential. To this end,  we use FL to aggregate these standalone models and obtain a global model. We validate the generated model using CDF of path loss. 

In the proposed approach, we use two generative NN models, both of which have a two-stage structure, i.e., link and path models \cite{geraci2}. In the first stage, an NN is used as a link model to determine the state of the link - whether it is LOS, NLOS, or no link, according to 3GPP requirements \cite{3gpprelease16}. To determine the link state, the relative position of the UAV to the gNB and the type of gNB are used as inputs. After the link state is determined, a generative model, i.e., a path model, is used in the second stage to generate the path parameters. This generative model is trained to match the distribution of the training dataset.
To perform the distributed training using FL, we trained the link-state model for each city and stored it on the corresponding station to use with the path model in FL. We then aggregate these generative models as described in Section \ref{sec:federatedVAE} and Section \ref{sec:federatedcgan}, respectively.
Once the model is trained, it can be used in simulations to statically determine channel parameters considering the link status.

\section{Federated VAE}
\label{sec:federatedVAE}
 In this section, we describe our FL-VAE for channel modeling of the path loss component. We first introduce the basic concepts of VAE to understand its content (Section \ref{vaeexp}) and then describe in detail the FL-VAE approach (Section \ref{FL-VAE}) used for modeling the channel parameters.
\subsection{Variational Autoencoder (VAE)}
\label{vaeexp}
VAE consists of encoder and decoder modules, where the encoder, defined as $q_{\theta}(z|x)$, characterizes the distribution of the input variables $x$ according to the encoding in the latent space $z$ (encoded representation of the input variables). On the other hand, the decoder, defined as $p_\phi (x|z)$, characterizes the distribution of the decoded variables based on the latent space, where $\theta$ and $\phi$ are the parameters of the encoder and decoder NNs, respectively.
The loss function of the VAE given in \cite{Kingma2013AutoEncodingVB} is as follows:

\begin{equation}
    \mathcal{L}(\phi,\theta) = -\mathbb{E}_{q_{\theta}(z|x_i)}\bigl[ \log p_\phi (x_i|z)\bigr] +  KL(q_{\theta}(z|x_i) \| p (z))
\end{equation}

The first component of the expression represents the reconstruction loss corresponding to the expected negative log likelihood of each data point. The expected value is calculated based on the encoder's distribution over the representations, and this component is intended to provide an incentive for the decoder to acquire the ability to reconstruct the data. The second term is the KL divergence, which acts as a regulizer and measures the loss of information when we use $q_{\theta}(z|x_i)$ to represent $p(z)$, which is the posterior distribution defined for the latent space, i.e., a Gaussian distribution.
\subsection{FL-VAE}
\label{FL-VAE}
FL-VAE uses the same VAE architecture proposed in \cite{geraci2} and trains generative (path) model using the FL framework developed in \cite{bano2022kafkafed}.
The goal of FL-VAE is to capture the conditional distribution $p(x|u)$ of all participating cities such that it tends to encode the local latent space of all cities into a single latent space and form the generic global model for generating channel parameters. VAEs can easily be trained in an FL framework since their encoder and decoder components comprise of NNs. Let $\mathcal{V}\coloneqq (\theta^e, \theta^d)$ be the VAE parameters, and $\theta^e$ and $\theta^d$ be the weights of the encoder and decoder, respectively. A centralized server initiates the training by communicating the initial weights of VAE $\mathcal{V}^t$ to all agents in the participating city stations. Each agent in a city initializes its own VAE model with these weights and uses local training data and a pre-trained link model to obtain a latent representation of its own data. Local updates of  each city $k$ is given by:

\begin{equation}\label{eq:local_grad}
    \mathcal{V}^{t+1}_k \longleftarrow \mathcal{V}^t_k - \eta \nabla  \mathcal{L}(\mathcal{V}^t_k)
\end{equation}

Where $\eta$ is the learning rate. Each city agent uses equation~(\ref{eq:local_grad}) to perform some local training epochs on local data and send the updates $\mathcal{V}^{t+1}_k$ to the central server. The server finally amalgamates the received updates with a weighted average approach given by:

\begin{equation}\label{eq:vae_aggr}
        \mathcal{V}^{t+1} = \sum_{k = 1}^{K} \frac{n_k}{n} \mathcal{V}^{t}_{k}
\end{equation}

$n_k$ are the number of training examples at each agent $k$ and $n$ is the total number of training data of each city. The server continues training until it obtains a global latent representation sufficient to represent all training data. 

\section{Federated CGAN}
\label{sec:federatedcgan}
In this section, we describe our FL-GAN approach to channel modeling. We first describe the Generative Adversarial Network (GAN) (Section \ref{ganexp}) and then the FL-GAN (Section \ref{FL-GAN}) used to model the channel parameters to form the generalised or universal model.
\subsection{Generative Adversarial Network (GAN)}
\label{ganexp}
The GAN is a popular concept first proposed in \cite{goodfellow2020generative}. Its main purpose is to generate synthetic data that closely resembles real data. GANs use an unsupervised learning approach to detect patterns in the input data and generate new samples with the same distribution as the original data. It consists of two NNs: the generator (G) and the discriminator (D), which compete in a "min-max two-player game." The G generates synthetic (fake) data from the learned latent vector, while the D discriminates the synthetic data from the real data. These models are trained until the G replicates the original data so well that it becomes difficult for the D to distinguish between the fake and the real data.

To generate samples from a given target, the CGAN was introduced in \cite{mirza2014conditional}. A CGAN learns the mapping from an observed sample $x$ and a random noise vector $z$ to an output sample $y$, represented as $G:{x,z}\xrightarrow{} y$, where $G$ is the generator function. Both networks in CGAN aim to solve a "min-max loss" like GAN given by \cite{mirza2014conditional}:
  \begin{multline}
     \mathcal{L}_{cGAN}(\mathcal{G},\mathcal{D}) =  \mathop{\mathbb{E}_{x,y}} \bigl[\log (\mathcal{D}(x,y))\bigr] + \\ \mathop{\mathbb{E}_{x,z}}\bigl[ 1 - \log (\mathcal{D}(x,\mathcal{G}(x,z)))\bigr]
 \end{multline}

G and D compete according to equation (2), where D tries to maximize the probability of assigning correct labels, and G tries to minimize that probability. In the next section, we describe the distributed approach using FL to train the CGAN.
\subsection{FL-GAN}
\label{FL-GAN}
We use the FL technique to train CGAN in a distributed manner. The training process is initiated by a central server, which communicates the initial parameters of generator and discriminator i.e., $\theta^G$ and $\theta^D$ to the agents in the cities. Each city agent initializes its own CGAN instance with the received parameters and trains it using local data and associated link state models. The updated parameters are then reported back to the server, which aggregates the updates from all cities as follows:

\begin{equation}\label{eq:aggr}
    \theta^G = \sum_{k = 1} ^{K} \frac{n_k}{n} \theta_k^G \quad ; \quad \theta^D = \sum_{k = 1} ^{K} \frac{n_k}{n} \theta_k^D
\end{equation}

$\theta^G$ and $\theta^D$ in equation (\ref{eq:aggr}) are the aggregate parameter estimates of G and D, respectively. The server goes through this process until it develops a global CGAN that can generate synthetic samples from the distribution that captures the local data distributions. After training, each local city unit can generate the path parameters with $\theta^G$. 

\section{Simulation Results}
\label{sec:simulationresults} 
In this section, we describe the performed experiments to assess the efficiency and effectiveness of the proposed FL approach.

\subsection{Dataset and settings}
In this work, we use raytracer data provided by \cite{geraci}. The dataset consists of channel parameters from five different cities, each with different landscapes and structures. However, for this work, we use the channel parameters of three cities (Beijing, London, and Boston).
In the raytracer simulation, the transmitting UAVs are positioned at different horizontal locations in each environment, with four possible heights: 30, 60, 90 and 120 m, to create the whole city dataset. 
A total of 36k links were created for Beijing, 25.8k for London, and 23k for Boston. 
All simulations were performed at a frequency of 28 GHz.

For our learning models, we used two generative NNs and trained them in a distributed manner using FL to make FL-VAE and FL-GAN model. The main goal is to develop a distributed model using FL framework that can be used universally for estimating channel parameters. In this context, we compare the generative models trained in a distributed manner and analyse which model is better at capturing channel characteristics of different latent spaces. We compared the results of these distributed models with the basic stand-alone models trained for each city using different statistical parameters, i.e., KL divergence and Wessterstein distance. The architecture and hyperparameters used to train these models are shown in Table \ref{T:model_summary} and Table \ref{T:hyp_setting} respectively. 

As mentioned earlier, in all cases our generative models consist of two cascaded models, the first of which is the link predictor and the second is the path generator. We first train the link predictor for each city separately and then use these pre-trained link models for simulation. 

\subsection{Results}
In this work, we propose a promising solution for extending the channel model to large-scale application scenarios by using a cooperative modeling approach with multiple distributed channel datasets. 
We first describe the results obtained in both centralized and distributed approaches. To ensure a fair comparison, we train all models with the same number of epochs and hyperparameters. In particular, we train the stand-alone models for 500 epochs and for the FL-VAE and FL-GAN models, we perform 100 rounds of local training, where each city trains its respective model for 5 epochs on its local data within each FL round.
\subsubsection{Stand-alone Models}
Our goal is to measure the extent to which the data generated by the generative models (VAE and GAN networks) are comparable to the test data. To this end, we compare the CDF of the path losses of the generated and test data. Both trained generative models are able to capture the dual-slope nature of CDF, which is a crucial component for the effectiveness of our proposed framework. However, due to space constraints, we only show in Table \ref{T:data_model_dist} the distance between the distribution of the standalone models (VAE and GAN) and the distributed models, i.e., (FL-VAE) and (FL-GAN).
\subsubsection{FL-VAE}
\begin{figure*}[ht]   
\subfloat[Beijing]{\includegraphics[width=0.3\textwidth, keepaspectratio]{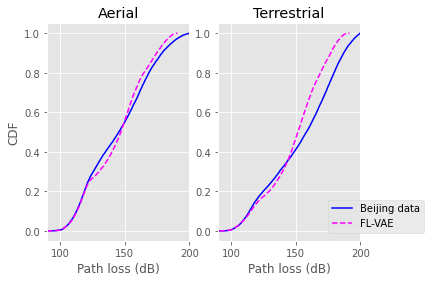}\label{fig:VAE_Beijing}}
\hspace{\fill}
\subfloat[Boston]{\includegraphics[width=0.3\textwidth, keepaspectratio]{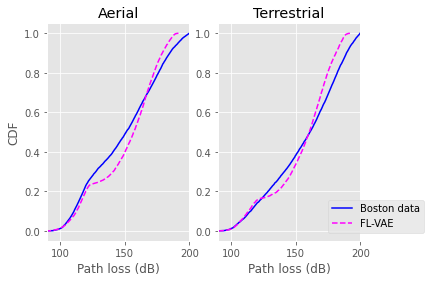}\label{fig:VAE_Boston}}
\hspace{\fill}
\subfloat[London]{\includegraphics[width=0.3\textwidth, keepaspectratio]{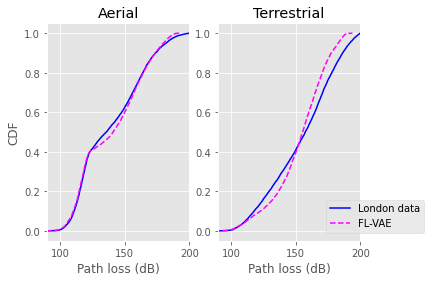}\label{fig:VAE_London}}

\caption[]{Federated Model (FL-VAE)}
\label{fig:fl-vae}
\end{figure*}

To evaluate the performance of our proposed decentralized model, we created CDF plots for the path losses of both the test data and the path losses generated by the FL-VAE model for each city. This allowed us to evaluate the generalizability of our federated global model, particularly in terms of its ability to accurately capture the channel characteristics of all participating cities. The results in Figures \ref{fig:VAE_Beijing}, \ref{fig:VAE_Boston}, and \ref{fig:VAE_London} show that our federated model performs better compared to the individual models of each city. 
In addition, the FL-VAE approach helps address potential privacy and security issues related to data sharing between different cities. These measures ensure that individual city data sets are not shared outside of the city, thus maintaining privacy and security. 
\subsubsection{FL-GAN}
\begin{figure*}[ht]   
\subfloat[Beijing]{\includegraphics[width=0.3\textwidth, keepaspectratio]{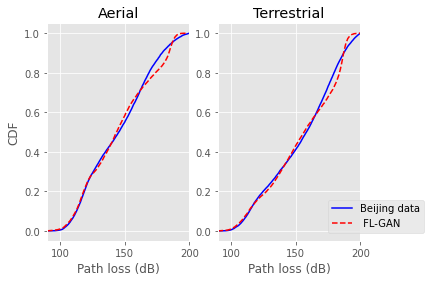}\label{fig:CGAN_Beijing}}
\hspace{\fill}
\subfloat[Boston]{\includegraphics[width=0.3\textwidth, keepaspectratio]{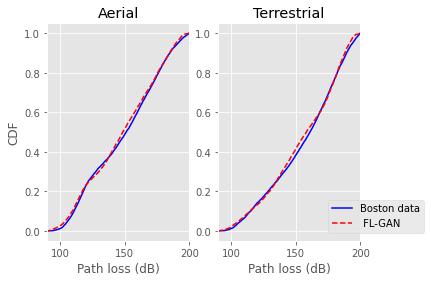}\label{fig:CGAN_Boston}}
\hspace{\fill}
\subfloat[London]{\includegraphics[width=0.3\textwidth, keepaspectratio]{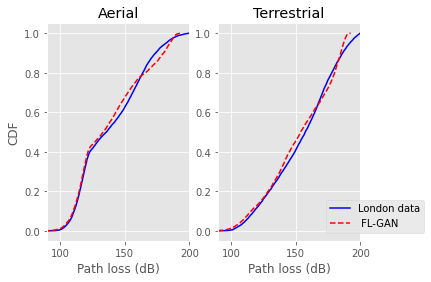}\label{fig:CGAN_London}}

\caption[]{Federated Model (FL-GAN)}
\label{fig:fl-gan}
\end{figure*}

Now we use the CGAN instead of the VAE to generate the channel parameters and compare its performance with the results we obtained with the FL-VAE and standalone GAN models. Our results show that the generative network learns the distribution of the channel modelling data very well and generates samples that exactly reflect the same distribution of the training dataset. It is also clear from Figures \ref{fig:CGAN_Beijing}, \ref{fig:CGAN_Boston}, and \ref{fig:CGAN_London} that FL-GAN produces better results for the path loss component of the channel parameters compared to FL-VAE. The results show that the channel parameters reconstructed using the FL-GAN approach are closest to the original test data and outperform the VAE-based methods. This can be attributed to the fact that it is difficult for VAEs to encode heterogeneous datasets from different cities into a common latent space, while GANs are better at learning diverse data. The FL-GAN approach is therefore better suited to deal with the challenges of heterogeneous data and produce synthetic data that accurately represents the actual data distribution.

\subsection{Performance Metrics and Evaluation Results}
This section presents the evaluation metrics used to assess the performance of the proposed distributed techniques FL-VAE and FL-GAN in generating synthetic data compared to the standalone models trained separately for each city. Table \ref{T:data_model_dist} summarizes the KL divergence and Wasserstein distance results obtained by comparing the test data distribution with the synthetic data distribution generated by the VAE, GAN, FL-VAE, and FL-GAN networks. These metrics are used to measure the distance between the test data distribution and the synthetic data generated by each model, and provide information about the accuracy and quality of the generated data.
These evaluation metrics used in Table \ref{T:data_model_dist} show that the distribution of synthetic data generated by the FL-GAN network is much closer to the true distribution compared to the other methods, i.e., the standalone networks and FL-VAE. This highlights the superiority of the proposed approach FL-GAN in accurately modeling the data and generating synthetic data that is very similar to the real data.

 As shown in Table~\ref{T:data_model_dist}, the KL-divergence between the test data distribution and the synthetic data distribution of the standalone GAN -model is much higher than that of the other alternatives. FL-GAN achieves the lowest KL -divergence among the alternatives, which is due to the fact that GANs generally require more training time than VAEs, but can generate better samples. We also evaluate our method using Wasserstein distance, which considers metric space. Table~\ref{T:data_model_dist} shows that FL-GAN significantly outperforms all other methods and achieves satisfactory performance in developing a global model for channel estimation parameters.

\begin{table}[t]
\caption{Model summary of the link model, path model (VAE and CGAN)}
\label{T:model_summary}
\begin{center}
\resizebox{\columnwidth}{!}{\begin{tabular}{c  c  c  c c }
\hline

\multirow{2}{*}{ \textbf{Model}} & \textbf{Number of} & \textbf{Hidden} & \textbf{Number of} & \textbf{Number of}\\
& \textbf{Inputs} & \textbf{Units} & \textbf{Outputs} & \textbf{Parameters} \\
\hline
Link Model & $5$ & $[25,10]$ & $3$ & $1,653$\\
VAE (Enc) & $125$ &$[200,80]$ & $40$ & $44,520$\\
VAE (Dec) & $25$ &$[80,200]$ & $240$ & $40,720$\\
GAN (Disc) & $125$ & $[1120,560,280]$ & $40$ & $1,055,761$\\
GAN (Gen) & $25$ &$[280,560,1120]$ & $240$ & $1,094,360$\\
\hline
\end{tabular}}
\end{center}
\end{table}

\begin{table}[t]
\caption{Hyperparameter settings for link model, Federated VAE and CGAN models}
\label{T:hyp_setting}
\begin{center}

\resizebox{\columnwidth}{!}{\begin{tabular}{ c  c  c  c }
\hline
\textbf{Item} 
& \textbf{Link Model} & \textbf{Generative Model (VAE)} &   \textbf{Generative Model (GAN)}\\
\hline
Communication Rounds & $N/A$ & $100$ & $100$\\
Epochs & $30$ & $5$ & $5$\\
Batch Size & $100$ & $100$& $100$\\
Learning Rate& $10^{-3}$ & $10^{-4}$ & $10^{-4}$\\
Optimizer & Adam & Adam &Adam\\
\hline
\end{tabular}}
\end{center}
\end{table}

\begin{table}[t]
\caption{Distance between real and generated data distribution for channel modeling parameters, the bold letters represents the smaller distance}
\label{T:data_model_dist}
\begin{center}
\resizebox{\columnwidth}{!}{\begin{tabular}{ l c  c  c}
\hline

\textbf{City} & \textbf{Method} & \textbf{KL-Divergence} & \textbf{Wasserstein Distance} \\
\hline

\multirow{4}{*}{Beijing} & VAE & $1.91$ & $13.92$ \\
& GAN & $3.08$ &$13.09$ \\
& FL-VAE & $1.63$ & $13.55$ \\
& FL-GAN &  \textbf{1.51} & \textbf{12.47} \\
\hline
\multirow{4}{*}{Boston} & VAE & $2.35$ & $12.48$ \\
& GAN & $1.66$ &$11.63$ \\
& FL-VAE & $2.29$ & $12.05$ \\
& FL-GAN & \textbf{1.25} & \textbf{11.33}  \\
\hline
\multirow{4}{*}{London} & VAE &  $1.70$ & $14.03$  \\
& GAN & $3.29$ &$12.86$ \\
& FL-VAE & $1.69$ & $13.95$  \\
& FL-GAN & \textbf{1.25} & \textbf{12.50}  \\
\hline

\end{tabular}}
\end{center}
\end{table}

\section{Conclusion}
\label{sec:conclusions}
NTNs are anticipated to play a crucial role in future wireless networks due to their cost efficiency and wide coverage area.
In this paper, we present a comprehensive study that employs a generative framework based on NNs to model wireless channels in a distributed environment. In order to have a common model for different cities, we train distributed generative models and combine them into a unified and adaptable model. Specifically, we propose a channel model for air-to-ground communication of UAVs in mmWave frequency bands. Our distributed training method does not require any special knowledge or technical expertise, as it learns directly from massive raw channel data to develop a generic channel model. The use of generative NNs, especially GANs and VAEs, is a suitable method for statistical channel modeling in complex scenarios. Although both models are capable of capturing data dependencies, our results show that the proposed FL-GAN approach outperforms the FL-VAE and centralized baseline methods in terms of learning the accuracy of path loss parameters. We validate our results with various statistical parameters, and the resulting model shows effective learning and interesting non-obvious predictions.


\bibliographystyle{IEEEtran}
\bibliography{bibliography/bibliography}

\end{document}